\documentclass[prb,preprint,amsmath,amssymb,showpacs,floatfix,superscriptaddress]{revtex4}

\usepackage{graphicx}
\usepackage{dcolumn}
\usepackage{bm}
\pdfoutput=1

\begin{document}
\title{Magneto-optical readout of dark exciton distribution in cuprous oxide.}
\author{D. Fishman}
\affiliation{Zernike Institute
for Advanced Materials, University of Groningen, 9747 AG Groningen,
The Netherlands.}
\author{C. Faugeras}
\author{M. Potemski}
\affiliation{Grenoble High Magnetic Field Laboratory, Centre National de la Recherche Scientifique, 25 Av. des Martyrs, 38
042 Grenoble, France}
\author{A. Revcolevschi}
\affiliation{Institut de Chimie Mol\'{e}culaire et des Materiaux d'Orsay - UMR
8182, Universit\'{e} de Paris Sud, B\^{a}timent 410, 91405 Orsay Cedex, France}
\author{P.H.M. van Loosdrecht}
\email{P.H.M.van.Loosdrecht@rug.nl}
\affiliation{Zernike Institute
for Advanced Materials, University of Groningen, 9747 AG Groningen,
The Netherlands.}

\date{\today}

\begin{abstract}
An experimental study of the yellow exciton series
in Cu$_2$O in strong magnetic fields up to 32 T shows
the optical activation of direct and phonon-assisted paraexciton
luminescence due to mixing with the quadruple allowed orthoexciton state.
The observed phonon-assisted luminescence yields information on the
statistical distribution of occupied states.
Additional time-resolved experiments provide a unique opportunity to directly
determine the time evolution of the thermodynamical properties of
the paraexciton gas. Because the lifetime of
paraexciton is hardly affected by the optical activation in
a strong magnetic field, this opens new possibilities for studies
aiming at Bose-Einstein condensation of excitons in bulk semiconductors.
\end{abstract}

\maketitle

\section{Introduction}

There are a number of well established phenomena which evidence that composite bosons made of an even number of fermions, such as for example He-4 atoms, Cooper pairs, or alkali-metal atoms, behave under some conditions like ideal bosons which can condense into a macroscopic quantum state known as the Bose-Einstein condensate (BEC).
One of the simpliest forms of such a boson is the exciton, {\em i.e.} a bound electron-hole pair which
characteristically appears as sharp emission and absorption features in the optical spectra of semiconductors and insulators.
Strong indications exist that Bose-Einstein condensation of excitons, predicted a long time ago \cite{keldysh65}, indeed occurs in man-made systems containing bilayer two-dimensional electron gases.\cite{eisen04}
The excitons in these systems are formed under equilibrium conditions.
The appearance of Bose-Einstein condensation of optically-created, finite-lifetime excitons is less evident \cite{snoke90, butov03} as only complex systems such as, for example, excitonic polaritons in semiconductor microcavities provide signatures of a condensate phase \cite{kasprzak06}.
The simplest system which has dominated the search for BEC in optically pumped solids is the excitonic gas in cuprous oxide, Cu$_2$O.\cite{snoke02}
Since excitons in Cu$_2$O are strongly bound, their bosonic character persists up to high densities and temperatures
and, due to the light masses of particles, their condensate form is expected at appealingly high temperatures.
Cuprous oxide exhibits a number of exciton series, the lowest of which is the so-called yellow series just below the
fundamental band gap, which by itself is again subdivided into the ortho- ($J=1$) and para-exciton ($J=0$)
series.\cite{snoke02}
Up to recently, research has mainly focused on the triply degenerate optically active orthoexciton gas in cuprous oxide\cite{snoke87, snoke90, den02, kar05, mos05}. It appears, however, that the lifetime of the orthoexciton, which is intrinsically limited by down-conversion to the lower
lying paraexciton state, is too short to reach the condensed state\cite{kar05}. In contrast, the lifetime for the paraexciton state, which is not optically active, is expected to be long. The exact value for the paraexciton lifetime varies widely from of nanoseconds to milliseconds \cite{mys79, den02, jan04, yosh07, fis06, kar05a} depending strongly on the quality of the sample, i.e., on the efficiency of nonradiative processes determined by impurity and defect concentration. 
Significant population of the paraexciton state can be achieved via laser pumping of higher-energy, optically-active states and the subsequent efficient relaxation processes toward the lowest singlet state. Advantageous on one side, the optical darkness of excitons at singlet state does not, however, allow to easily probe their properties and this has been one
of the main obstacles in searching of their possible condensate form.

Although direct optical absorption and emission to and from the
paraexciton ground state are strictly forbidden, paraexcitons can
radiatively decay with the assistance of $\Gamma_5$ phonon,
\cite{snoke90, suil99}. In this process the spin-flip is allowed by
spin-orbit-lattice interaction. Unfortunately, the radiation
efficiency is several orders of magnitude smaller than that of the
orthoexcitons,  requiring long integration time to detect the
emission. Therefore one has to rely on alternative methods to
study the properties of the paraexcitons. One of the methods pursued
so far has been to study the optically allowed inter-exciton
transitions (similar to the Lyman series in the hydrogen spectrum),
\cite{jol02, kuw04, kub05, kar05a, fis06, yosh07}. Although one can
easily extract basic parameters such as the effective mass and
the lifetime of paraexcitons, it is more difficult to gain
information on the statistical properties of the exciton gas from
these kind of experiments. A second method is
to break the symmetry, thereby optically activating direct and
phonon-assisted transitions to and from the paraexcitonic ground
state. There are many ways to break this symmetry, including
applying pressure \cite{lin93,nak03,nak03a} and applying electric
fields \cite{ett03}. These strong perturbations typically lead
to a considerably reduced lifetime of the paraexciton \cite{nak02}.
Application of a magnetic field breaks the symmetry in
a more subtle way. In this case, the optically forbidden paraexciton
state mixes with the quadruple allowed orthoexciton state, leading
to weak para-exciton emission, which intensity is quadratic in the
field, and a negligible influence on the lifetime.
early studies already uncovered  paraexciton magneto-luminescence
\cite{gas83,kono99}, although due to the relatively low magnetic
fields, and the chosen crystal orientation these experiments did not
allow for a detailed field dependent study nor for a study of the
statistical properties of the para-exciton gas. More recently, the
optically forbidden paraexcitons were observed in fields up to 10 T,
In this work the authors used the resonant excitation to create
an initially cold paraexciton gas. In addition, these authors
observed a surprisingly narrow para-exciton magneto-absorption line.
\cite{bran07, san08}.

In this paper we show that application of sufficiently high magnetic fields is a
non-invasive and subtle tool to activate both the direct and phonon-assisted
emission of the initially dark paraexcitons in Cu$_2$O.
The observation of phonon-assisted emission provides a direct tool to probe
the thermodynamical properties of paraexcitons gas in Cu$_2$O, and eventually the
appearance of its possible condensate phase.

\section{Experimental details.}

In the present work, the paraexciton magneto-luminescence has been
studied in a Faraday geometry with magnetic fields up to $B=32$~T at
a temperature of $T=2$~K. This geometry was chosen to optimize
both the degree of polarization, as well as the emission efficiency
\cite{ell61,yak84,kono99}. For the steady state experiments,
platelets (thickness 200~$\mu$m) cut and polished from floating zone
grown Cu$_2$O single crystals\cite{sch74} of [110] orientation
(denoted below as Sample I) were placed in a pumped liquid helium
bath cryostat (1.3 K) and excited with the 532 nm line of a 100 mW
solid state laser. The circular polarized magneto-luminescence
spectra\footnote{Luminescence spectra are only 85 \% polarized due
to the experimental conditions.} were detected using monochromator (resolution 0.02 nm) equipped with a LN$_2$ cooled CCD
camera. Since the paraexciton lifetime of Sample I was found to be less
than 30 ns, indicating the presence of impurities, we used a
[110] oriented 250~$\mu$m thick platelet of  cut and polished from a
high purity Cu$_2$O natural crystal (denoted below as Sample II)
for the time-resolved experiments.
A pulsed solid state laser (50 ns, 4
kHz repetition rate, 523 nm, 0.6 mJ/cm$^2$) was used as an excitation source.
The emitted light was detected with a single monochromator (resolution
0,06 nm)  equipped with a fast CCD detector, yielding the time resolution of 30 ns.

\begin{figure}[ht]
\begin{center}
\includegraphics[width=8cm]{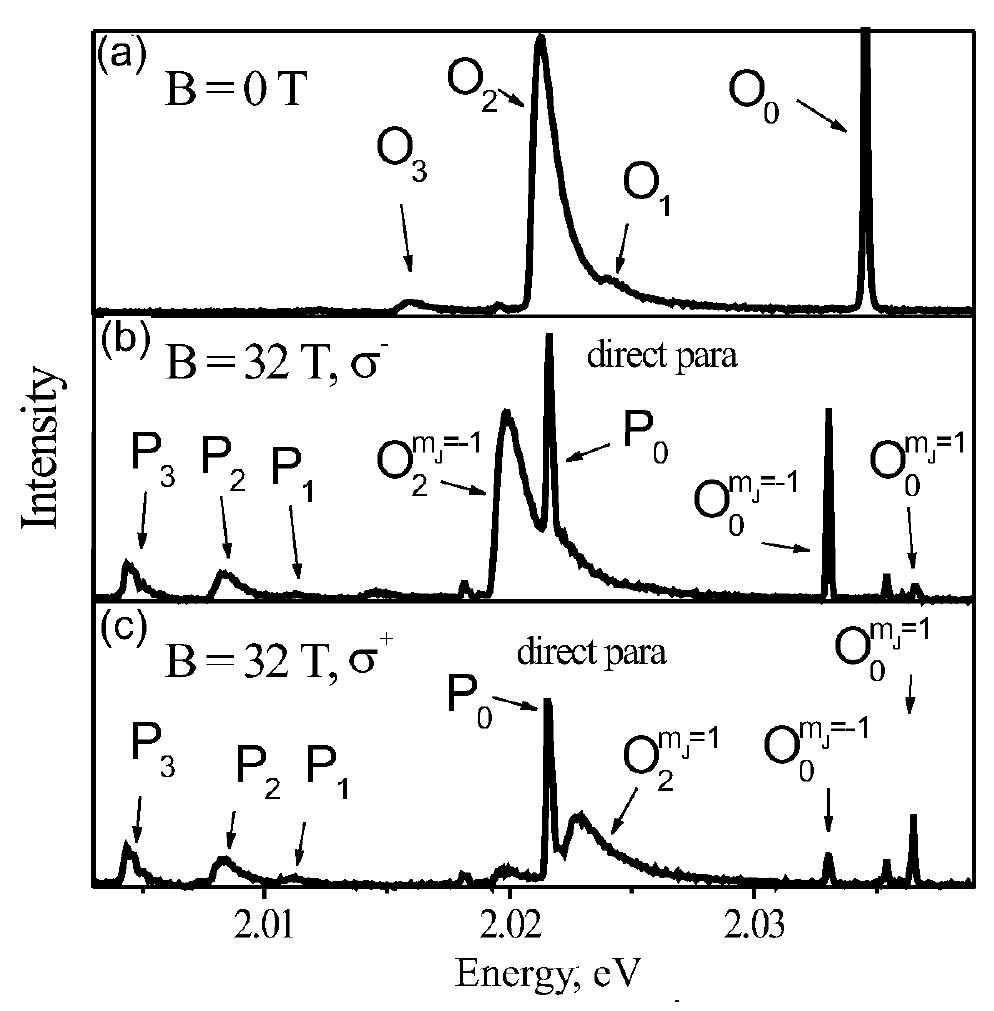}
\end{center}
\caption{
Exciton magneto-luminescence spectra of Sample I at bath temperature $T=2$~K.
a) Unpolarized luminescence at $B=0$~T.
O$_0$: Direct ortho-exciton emission;
O$_1$ - O$_3$: Phonon assisted ortho-exciton emission involving
$\Gamma_5$ (6.6 meV), $\Gamma_3$ (13.2 meV), and $\Gamma_4$ (18.2 meV) phonons,
respectively.
b) and c) Circularly polarized [b) $\sigma^+$, c) $\sigma^-$]
luminescence at $B=32$~T.
P$_0$: Direct para-exciton emission,
P$_1$-P$_3$: Phonon-assisted para-exciton emission involving
$\Gamma_5$, $\Gamma_3$, and $\Gamma_4$ phonons, respectively.
} \label{fig1}
\end{figure}

\section{Results and discussions.}

Figure~\ref{fig1} shows some typical luminescence spectra obtained
for $B=0$~T [a), unpolarized] and $B=32$~T [b), right, and c), left
circularly polarized]. As expected, the $B=0$~T luminescence
spectrum is dominated by orthoexcitonic features. Apart from the
quadrupole allowed direct orthoexciton transition ($O_0$ at 2.033
eV), there are clear lines from phonon-assisted transitions
involving $\Gamma_5^-$ ($O_1$ at 6.6 meV), $\Gamma_3^-$ ($O_2$ at
13.2 meV), and $\Gamma_4^-$ ($O_3$ at 18.2 meV) phonons
\cite{dev72}. In addition to these strong orthoexcitonic features, 
also a very weak emission originating from phonon 
assisted ($\Gamma_5^{-}$) paraexciton decay is observed at 2.011 eV.\cite{snoke90, suil99} 
The presence of a magnetic field leads to two
important changes in the spectra. First, the threefold degeneracy of
the triplet ortho-exciton state is lifted leading to a splitting of
the ortho-exciton derived bands (labeled O$_i^{m_J}$ in
figure~\ref{fig1}). The $\sigma^+$ and $\sigma^-$ spectra show
predominantly the $m_J=1$ and $m_J=-1$ components, respectively.
Second, and more interestingly for the present work, a number of 
emission features appear in the spectra, which are not polarized.
The sharp feature at 2.0216 eV originates from direct emission
from the para-exciton 1s to the ground state ($P_0$). In addition,
like for the ortho-exciton, three phonon-assisted paraexciton
recombination lines are observed at 2.011 eV (P$_1$, $\Gamma_5^-$
phonon), 2.083 eV (P$_2$, $\Gamma_3^-$) and 2.045 eV (P$_3$,
$\Gamma_4^-$ ), respectively. As is the case for orthoexcitons,
these latter emission lines may provide information on the statistical
properties of the optically excited paraexciton gas. Before turning
to these properties, we first discuss some of the more general
features of the para- and ortho-exciton emission.

\begin{figure}[hb]
\begin{center}
\includegraphics[width=8cm]{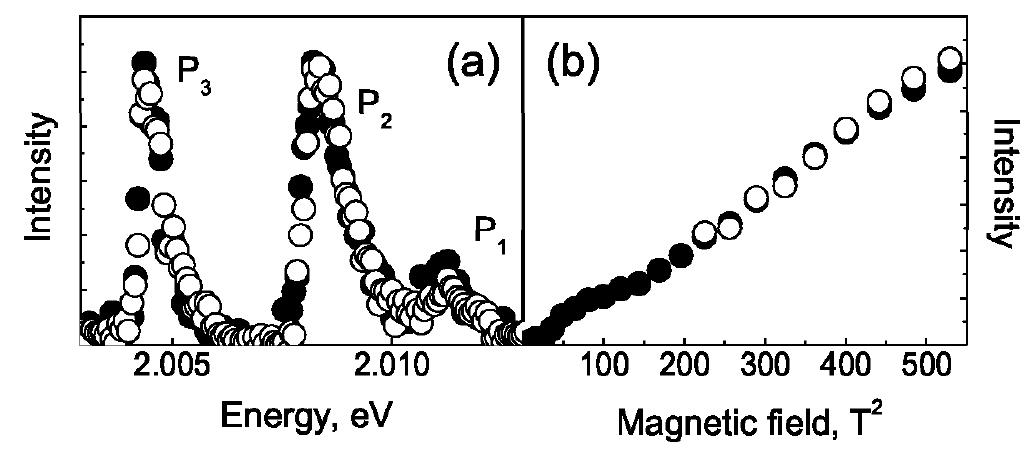}
\end{center}
\caption {(a) Right (open circles) and left (filled circles)
circularly polarized phonon assisted para-exciton emission at
$T=2$~K, $B=32$~T (Sample I). P$_1$-P$_3$ refer to $\Gamma_5$,
$\Gamma_3$, and $\Gamma_4$ phonon assisted transitions, respectively.
(b) Intensity of the $\Gamma_3$- phonon-assisted paraexciton
emission (closed circles) and direct paraexciton emission (open
circles) as a function of the squared magnetic field. \label{fig2}}
\end{figure}

The reason that the paraexciton lines are activated in a magnetic
field is the presence of the perturbing Zeeman term in the electron-hole
Hamiltonian:
\begin{equation}
V=\frac{1}{2}\mu_{B}(g_e S_{e}^{z}B_z+g_h S_{h}^{z}B_z),
 \label{equ1}
\end{equation}
leading to a mixing of the singlet para-exciton state $\left|S\right\rangle$
with the $m_J=0$ triplet ortho-exciton state $\left|T_0\right\rangle$.
The perturbed singlet state becomes
\begin{equation}
\left|S\right\rangle_B =
\left|S\right\rangle_0-\frac{(g_e-g_h)\mu_BB}{\Delta+\sqrt{\Delta^2+(g_e-g_h)^2\mu_B^2B^2}}
\left|T_0\right\rangle_0,
\label{equ2}
\end{equation}
where $g_e$ ($g_h$) is the electron (hole) $g$-factor, $\mu_B$ the
Bohr magneton, and $\Delta=12$~meV the energy difference between the
ortho- and paraexciton ground states. This mixing leads to a
non-zero para-exciton luminescence intensity, which in the limit
$(g_e-g_h)\mu_B B \ll \Delta$, as is the case in the experiments,
will be proportional to the square of the magnetic field. Moreover,
since the mixing occurs with the $m_J=0$ state only, one expects a linear polarization for the paraexciton emission. Experimentally,
these features are indeed observed, as is shown in Fig.~\ref{fig2}a)
and b): There is no difference in para-exciton emission intensity
for $\sigma^+$ and $\sigma^-$ polarized spectra (Fig.~\ref{fig2}a),
and the intensity is indeed quadratic in fields up to 32 T
(Fig.~\ref{fig2}b)

A second consequence of the mixing is that the energies of the
states involved show a diamagnetic shift. The field dependent energy
of the paraexciton state is given by
\begin{equation}
E_S(B) = E_S(0)+\frac{1}{2}\left(\Delta-\sqrt{\Delta^2+(g_e-g_h)^2\mu_B^2B^2}\right).
\label{equ3}
\end{equation}
In the same low field limit as before, this reduces to
$E_S(B)\approx E_S(0)-(g_e-g_h)^2\mu_B^2B^2/4\Delta$. Similarly, the
energy of the $m_J=0$ orthoexciton state is expected to shift to
higher energy by the same amount.

\begin{figure*}[htb]
\begin{center}
\includegraphics[width=14cm]{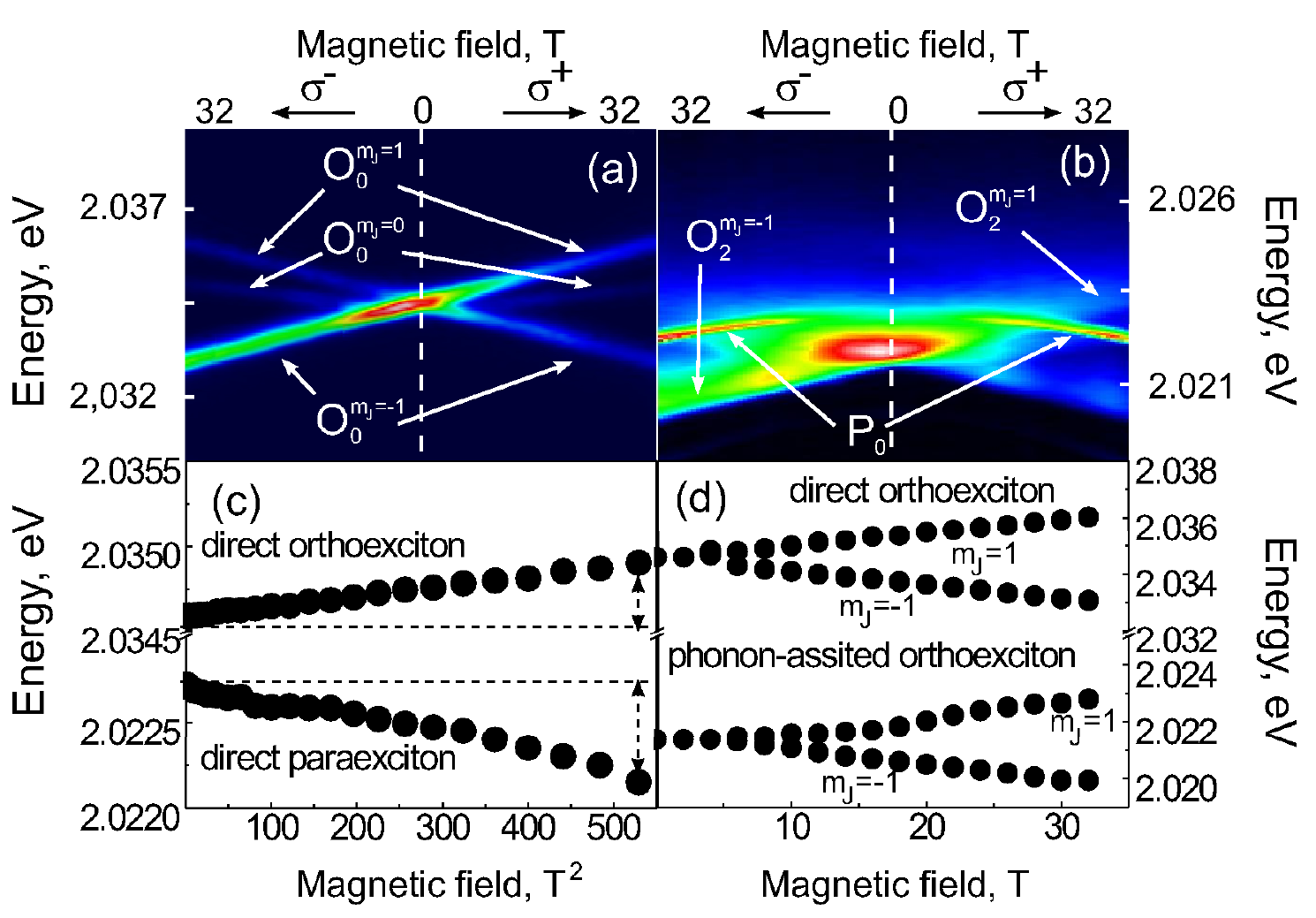}
\end{center}
\caption{ Two dimensional image of the field dependence of the 1s
orthoexciton spectra for Sample I at $T=2$~K in the energy range
of the direct emission (a) and the phonon-assisted emission (b). (c)
Energy of the direct paraexciton and m$_J$=0 orthoexciton emission
as a function of the magnetic field squared at $T=2$~K. (d) Energy
position of orthoexciton lines as a function of the magnetic field
at $T=2$~K. \label{fig3}}
\end{figure*}

These diamagnetic shifts are nicely demonstrated in Fig.~\ref{fig3},
which shows a two dimensional false color map of the ortho- and
para-exciton emission intensity as a function of energy and field.
In the presence of a magnetic field, the single $\Gamma_5^+$ ortho
emission line observed at $B=0$~T splits into its $m_J=-1,0,1$
components (Fig.~\ref{fig3}a). The same holds for the phonon
assisted ortho-exciton lines (Fig.~\ref{fig3}b). While the $m_J=1$
($\sigma^+$) and $m_J=-1$ ($\sigma^-$) lines show their expected
linear Zeeman splitting, the energy of the $m_J=0$ line indeed
increases quadratically with increasing field. In line with this,
the para-exciton line shows, once observable, a quadratically
deceasing energy upon increasing field (Fig.~\ref{fig3}b). However,
even though Eq.~\ref{equ3} predicts equal size shifts for the ortho-
and paraexcitons, the shift for the ortho-exciton at $B=32$~T (0.31
meV) is about half of the value observed for the paraexciton (0.57
meV), see Fig.~\ref{fig3}(c). The origin of this discrepancy is
presently not clear, although it has been suggested that it is due
to the interaction of the orthoexciton with higher lying
levels\cite{kuw77}.

In principle one can extract the values of the electron and hole
$g-$factors from the observed energy shifts, \cite{car66}. Assuming
that the paraexciton indeed shifts according to Eq.~\ref{equ3} leads
to the relation $(g_e-g_h)^2=7.5$. An independent relation can be
obtained from the observed Zeeman splitting, shown in more detail in
Fig.~\ref{fig2}(d), yielding $|g_e+g_h| = 1.66$. These values are in
good agreement with previously reported values \cite{car66,kuw77,
froh82, kon99}, and lead to either $g_e=2.68$, $g_h=-1.02$, or the
reverse.

\begin{figure}[htb]
\begin{center}
\includegraphics[width=8cm]{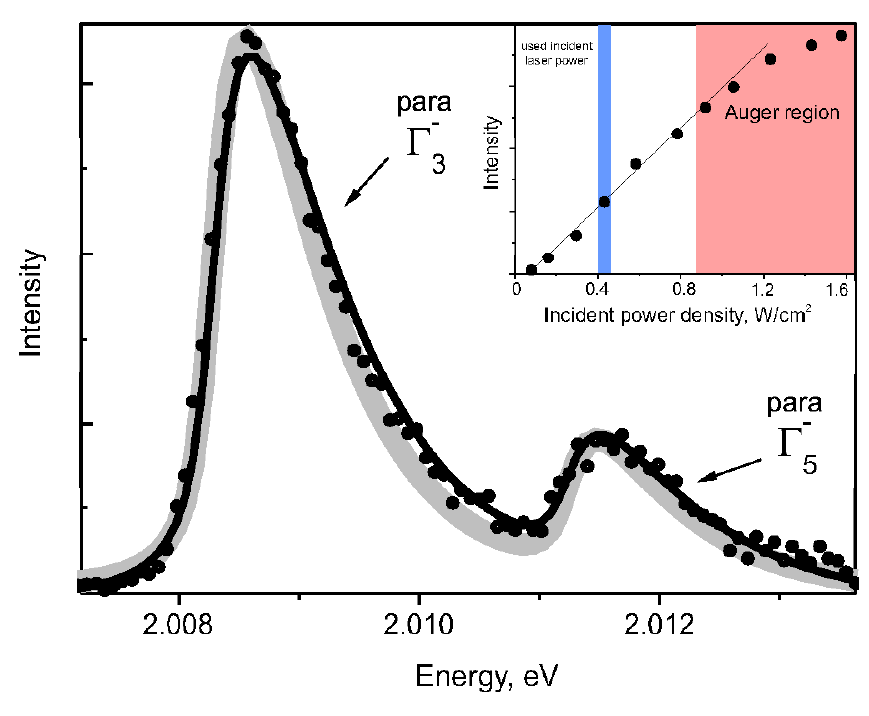}
\end{center}
\caption{ Phonon-assisted paraexciton emission at $T=2$~K,
$B=23$~T. Open circles - experimental data; grey line - theoretical
model using Maxwell-Boltzman distribution ($T=7.3$~K); black line -
theoretical model using Bose-Einstein distribution ($T=8$~K,
$\mu=-0.5$~meV ). Inset: The integrated intensity of phonon-assisted
luminescence as a function of the excitation power density.}
\label{fig4}
\end{figure}

Finally, we return to the phonon-assisted paraexciton lines
and the statistical properties of the paraexciton gas.
Fig.~\ref{fig4} (symbols)
shows the experimental $\Gamma_3^-$ and $\Gamma_5^-$
phonon-assisted lines at $T=2$~K for $B=32$~T.
Assuming energy independent exciton-phonon coupling matrix
elements and dispersionless phonons with energy $\hbar\Omega$,
the emission at a certain energy is given by \cite{kar05, snoke90},
\begin{equation}\label{equ5}
I(E)\propto \sum_i D(E-E_0-\hbar\Omega_i) f(E-E_0-\hbar\Omega_i)
\end{equation}
where $i$ labels the phonons, $E_0$ is the energy of the
${\mathbf k}=0$ paraexciton,
$D(E)\propto\sqrt{E}$ is the density of excitonic states,
and $f(E)$ the statistical distribution function for the
excitons.

\begin{figure*}[htb]
\begin{center}
\includegraphics[width=15cm]{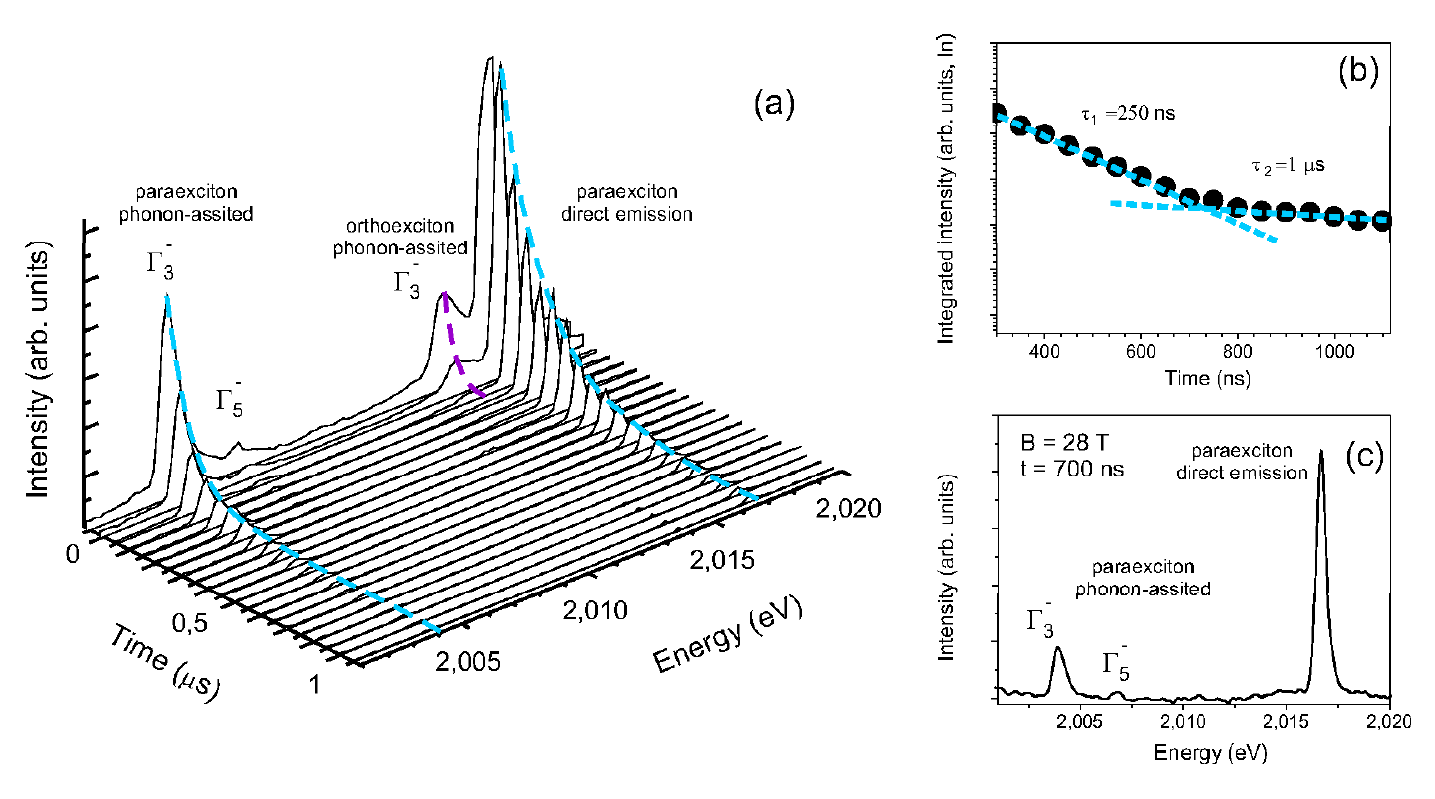}
\end{center}
\caption{(a) Exciton luminescence at different times after the
excitation with 50 ns pulse of 523 nm wavelength at $T=2$~K,
$B=28$~T for Sample II (pump energy density of 0.6 mJ/cm$^2$) (b) The
magneto-luminescence spectrum of the paraexciton gas at 600ns after
the excitation for Sample II.} \label{fig5}
\end{figure*}

\begin{figure}[htb]
\begin{center}
\includegraphics[width=7cm]{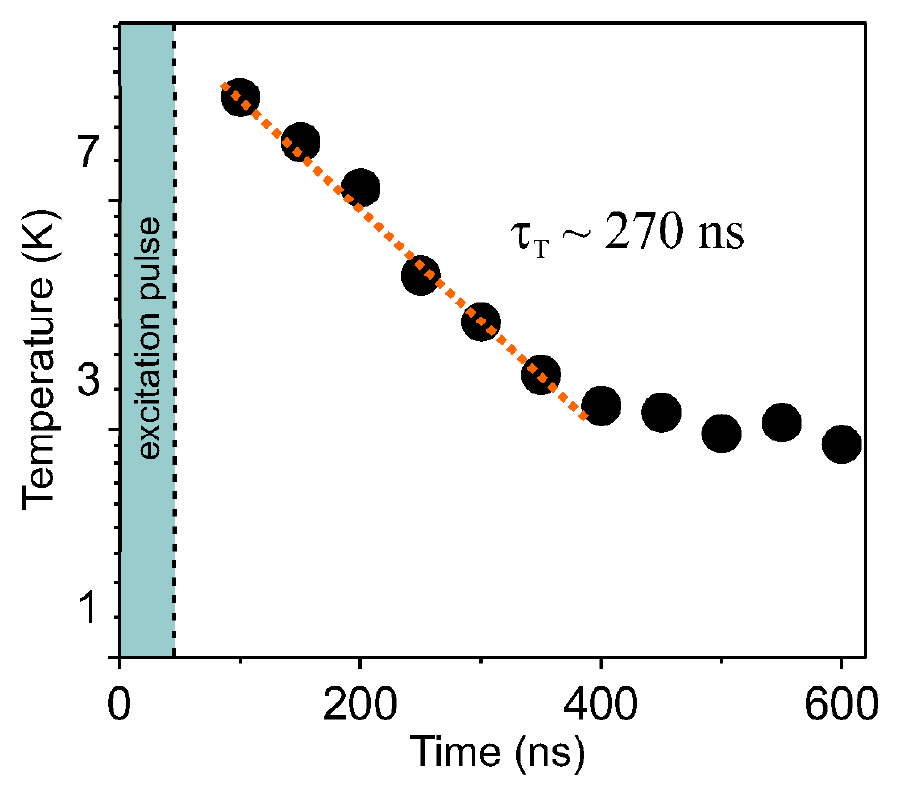}
\end{center}
\caption{Time evolution of the paraexciton gas temperature $T_{MB}$.
Excitation with 50 ns pulse of 523 nm wavelength. Bath temperature
$T=2$~K, Sample II.} \label{fig6}
\end{figure}

The experimental spectrum is given by convolution of Eq.~\ref{equ5}
with the known experimental resolution (0.1 meV). The drawn light
grey line in Fig.~\ref{fig4} represents a fit of the convoluted
Eq.~\ref{equ5} to the data using the classical Maxwell-Boltzmann
distribution. The parameter extracted from the fit is an exciton gas
temperature $T_{MB}=7.3$~K. One important parameter one would like
to extract from the experiment is the density of excited
paraexcitons. A straightforward, though inaccurate, method is to
assume that  each absorbed photon creates exactly one exciton, which
is then allowed to diffuse throughout the sample volume. Assuming that 50\% of the incident photons were absorbed, for the
current experimental conditions and given parameters, as penetration depth of 100 $\mu$m \cite{kar05}, this yields a upper limit for the
density of $\sim$10$^{16}$ cm$^{-3}$. In principle, one could
analyze the luminescence line shape using the Bose-Einstein
distribution, i.e. assuming quantum statistics, and from the
extracted gas temperature and chemical potential calculate the
density. The legitimacy of this analysis, in particular use of
Bose-Einstein distribution function, has, however, been questioned
in \cite{oha00}. It is suggested, that Auger processes might lead to
an overestimation of the exciton density and that the competition
between Auger recombination and cooling of the phonon-emission could
possibly result in a Bose-like 'distribution'. To avoid this
problem, the current experiments have been performed at moderate
excitation densities, where Auger processes are playing only a minor
role. This is illustrated in the inset of Fig.~\ref{fig4} which
shows the excitation power dependence of the intensity of the
phonon-assisted paraexciton emission at $B=23$~T. For low powers the
emission scales linearly with the excitation power. The influence of
Auger processes is observed only for excitation power densities
exceeding 0.8 W/cm$^2$, where the intensity shows the typical
sub-linear behavior with increasing power. Similar results were
observed in \cite{tra86, jang06} for the orthoexciton luminescence:
the intensity increases by the square root of the excitation
density. In present experiments, the used incident power had a value
well below 'Auger regime' (0.4 W/cm$^2$) ensuring that Auger process
play a minor role only.

A convolution of Eq.4 using quantum statistics has also been fitted to the data presented in
Fig.~\ref{fig4} (black curve). The parameter extracted from the fit are an exciton gas temperature $T_{BE}=8$~K and
a chemical potential $\mu=-0.5$~meV.
For the exciton gas to be truly in the quantum limit, the density has to be
higher than or at least comparable to the quantum density $n_Q=\left(\frac{mk_BT}{2\pi \hbar^2}\right)^{3/2}$.
For a para-exciton gas at 8 K this means a density of $\sim2\times10^{18}$~cm$^{-3}$.
Using the obtained chemical potential and temperature extracted from the fit, the calculated density of
the exciton gas gives an upper estimate of $10^{15}$~cm$^{-3}$, {\em i.e.} much smaller than the quantum density.
Therefore, the conclusion for the present experiments is that the exciton gas is still in the classical limit,
and one has to go to much lower temperature or higher densities to reach the quantum limit.

In addition to the above described continuous wave experiments, we
have also performed time resolved experiments to elucidate the
dynamics of the paraexciton gas. Sample II was used for these
experiments since it shows a relatively long paraexciton lifetime.
Fig. ~\ref{fig5} (a) shows some typical time-resolved
magneto-luminescence spectra at $T=2$~K, $B=28$~T. At early times,
the spectrum shows the clear presence of both ortho- and
paraexcitons. As expected, the orthoexcitonic features vanish
rapidly due to ortho-to-paraexciton conversion. In contrast, the
paraexcitons have a fairly long lifetime even at these high fields.
The observed decay is clearly not mono-exponential, as demonstrated
in Fig.~\ref{fig5}(b). It reveals a fairly complicated behavior.
There is a fast initial decay which closely follows the pump pulse
profile (60 ns). Directly after the pump pulse, a fast decay occurs
with a time constant of $\sim$250 ns leading to a reduction of the
density by about two orders of magnitude within the first 600 ns.
This decay is most likely due to the collision-induced
non-radiative decay process.\cite{yosh07} This is in line with the
high power density used in this experiment. A simple estimate yields
an upper limit for the initial density of about 10$^{17}$ cm$^{-3}$,
which is well within the auger regime. Finally after 600 ns auger
processes do not play an essential role anymore (the density dropped
below 10$^{15}$ cm$^{-3}$), and the luminescence intensity slowly
decays further with a time constant of around 1$\mu$s, mainly
limited by the purity of the sample.

Using the method described above , one can again extract an estimate of the
gas temperature (using a classical Maxwell-Boltzmann distribution function)
as a function of time after excitation.
The evolution of the estimated paraexciton gas temperature presented in Fig. ~\ref{fig6} shows
that there is a fast initial cooling with a time constant of 270 ns. This decay time is comparable
to that observed for the intensity decay, suggesting that also for the cooling Auger processes play
a dominant role. After about 500 ns the gas has cooled down to about 2 K. This temperature
corresponds to the bath temperature, meaning that the gas reachs thermodynamical equiibrium at these times.

\section{Conclusion}

In conclusion, we have studied the magneto-optical properties of the
yellow exciton series in cuprous oxide in magnetic fields up to
32T. We observed the direct emission of the normally optically inactive paraexciton
arising from mixing of the 1s paraexciton state with the 1s $m_J=0$ orthoexciton state. The
experimental results can be well explained using first order
perturbation theory.
Due to the gentle breaking of the symmetry by the magnetic field, the
lifetime of excitons is only weakly affected allowing efficient
thermalization of the exciton gas.
Besides the direct emission of the paraexciton, we also
observe 3 additional lines which we are assigned to
phonon-assisted emission processes from the paraexciton state,
involving various phonons of different
symmetries, similar to the orthoexciton case.
These lines provide a unique opportunity for direct
determination of the thermodynamical properties of the paraexciton gas.
For the present experiments, it is clear that the density of the exciton gas is well below the quantum density.
The time evolution of the gas parameters, obtained from time-resolved experiments are calling for further
experimental efforts at higher exciton densities and lower temperatures, preferably in a confined environment to avoid diffusion, 
to solve the enigma of BEC in Cu$_2$O.

We gratefully acknowledge Makoto Kuwata-Gonokami for fruitful discussions and for providing sample II. 
We also thank Foppe de Haan (University of Groningen) for assistance in data treatment. 
This work was partially supported by the European Access Program RITA-CT-2003-505474 and the
Grenoble High Magnetic Fields Laboratory (CNRS, France).

%\bibliography{bibliography}

\end{document}